\documentclass[prd,aps,preprint,amsmath,amssymb,eshowkeys,nofootinbib]{revtex4-1}
\pdfoutput=1
\usepackage{verbatim,graphics,graphicx,color,slashed,textcomp,bbm,mathdots,multirow,array}
\usepackage[normalem]{ulem} 
\usepackage[colorlinks=true,linkcolor=red,urlcolor=blue,citecolor=blue]{hyperref}

\graphicspath{{figures/}}

\usepackage[utf8]{inputenc}

\begin{document}

\title{Inflation in Weyl Scaling Invariant Gravity with $R^3$ Extensions}

\author{Qing-Yang Wang$^{a}$}
\author{Yong Tang$^{a,b,c,d}$}
\author{Yue-Liang Wu$^{a,b,c,e}$}
\affiliation{\begin{footnotesize}
		${}^a$University of Chinese Academy of Sciences (UCAS), Beijing 100049, China\\
		${}^b$School of Fundamental Physics and Mathematical Sciences, \\
		Hangzhou Institute for Advanced Study, UCAS, Hangzhou 310024, China \\
		${}^c$International Center for Theoretical Physics Asia-Pacific, Beijing/Hangzhou, China \\
		${}^d$National Astronomical Observatories, Chinese Academy of Sciences, Beijing 100101, China\\
		${}^e$Institute of Theoretical Physics, Chinese Academy of Sciences, Beijing 100190, China
		\end{footnotesize}}
	
\date{\today}

\begin{abstract}
The cosmological observations of cosmic microwave background and large-scale structure indicate that our universe has a nearly scaling invariant power spectrum of the primordial perturbation. However, the exact origin for this primordial spectrum is still unclear. Here, we propose the Weyl scaling invariant $R^2+R^3$ gravity that gives rise to inflation that is responsible for the primordial perturbation in the early universe. We develop both analytic and numerical treatments on inflationary observables, and find this model gives a distinctive scalar potential that can support two different patterns of inflation. The first one is similar to that occurs in the pure $R^2$ model, but with a wide range of tensor-to-scalar ratio $r$ from $\mathcal O(10^{-4})$ to $\mathcal O(10^{-2})$. The other one is a new situation with not only slow-roll inflation but also a short stage of oscillation-induced accelerating expansion. Both patterns of inflation have viable parameter spaces that can be probed by future experiments on cosmic microwave background and primordial gravitational waves.
\end{abstract}

\maketitle
\newpage

\section{Introduction}

Inflation is a hypothetical epoch of exponential expansion introduced in the very early universe to solve the cosmological horizon and flatness problems \cite{Guth:1980zm,Linde:1981mu}. It is also a reasonable scheme to explain the origin of primordial density perturbations, which plays the role of the seeds that formed the structure of current universe \cite{Mukhanov:1990me}. In recent years, the precise measurement of cosmic microwave background (CMB) presents us with an almost scale invariant spectrum of primordial perturbations \cite{Planck:2018jri}. This result is usually explained by an approximate de Sitter spacetime of the very early universe \cite{Mukhanov:1981xt, Hawking:1982cz, Guth:1982ec, Starobinsky:1982ee, Bardeen:1983qw}. Moreover, it is theoretically explored that there is a more profound and basic principle behind the phenomenon, namely, local Weyl scaling invariance of the universe. This symmetry is first proposed by H. Weyl in the attempt of understanding gravity and electromagnetism in a unified framework \cite{Weyl:1918ib, Weyl:1919fi}, and after a century of development, it has been applied extensively to particle physics, cosmology~\cite{Smolin:1979uz, Cheng:1988zx, Nishino:2009in, Romero:2012hs, Bars:2013yba, Quiros:2014hua, Scholz:2014kba, Ohanian:2015wva, Bamba:2015uxa, Ferreira:2016wem, deCesare:2016mml, Ferreira:2018itt, Ferreira:2018qss, Tang:2018mhn, Ghilencea:2018thl, Wetterich:2019qzx, Tang:2019uex, Tang:2019olx, Ghilencea:2021lpa, Ghilencea:2021jjl} and gauge theory of gravity~\cite{Wu:2015wwa, Wu:2017urh, Wu:2021ign, Wu:2021ucc}. 

Lately, inflation in the Weyl scaling invariant theory of gravity, especially induced by a quadratic curvature term $R^2$, has been of many concern \cite{Ghilencea:2018dqd, Ferreira:2019zzx, Ghilencea:2019rqj, Oda:2020yyv, Ghilencea:2020piz, Oda:2020hvu, Tang:2020ovf, Oda:2020cmi, Ghilencea:2020rxc, Cai:2021png, Wang:2022ojc}. Comparing with the conventional $R^2$ model, which is also called the Starobinsky model \cite{Starobinsky:1980te, Vilenkin:1985md, Mijic:1986iv, Maeda:1987xf}, the scaling invariant version not only allows a viable inflation scenario with good observational agreement, but also provides a framework to comprehend another fundamental puzzles, such as the hierarchy problem~\cite{Ghilencea:2019rqj, Oda:2020hvu, Aoki:2021skm} and dark matter candidates~\cite{Tang:2020ovf, Wang:2022ojc}.

However, inflation with only quadratic scalar curvature might be just a simplistic scenario. From the viewpoint of effective field theory, any higher-order curvature effects may exist and play a role in the early universe. Hence it is reasonable to evaluate their impacts on inflation. Generally, the extensions with high-order tensors, like $R_{\mu\nu}R^{\mu\nu}$ or $R_{\mu\nu\rho\sigma}R^{\mu\nu\rho\sigma}$, can result in unacceptable ghost degrees of freedom~\cite{Stelle:1976gc}, while the terms of arbitrary functions of the Ricci scalar are known to be safe. Therefore, in this paper, we consider a minimal extension of Ricci scalar beyond the $R^2$ model with Weyl scaling invariance, namely a cubic term coupled with an extra scalar field as denominator $R^3/\varphi^2$. We will show that even if this term is extremely small, it will have an essential impact on inflation, which even open up a completely different inflationary scenario from Weyl $R^2$ and conventional $R^2+R^3$ models.

The paper is organized as follows. In Sec.~II, we develop the analytic formalism of Weyl $R^2+R^3$ model and derive the effective scalar potential. We show that in some cases, the potential has two different kinds of global minima,  leading to two distinctive inflationary patterns. In Sec.~III, we investigate the inflation in the pattern of evolving to the side minimum. We calculate the spectral index $n_s$ and tensor-to-scalar ratio $r$ of the inflationary perturbations, and give the preferred parameter space allowed by the latest observations. Analytical treatments are developed for more transparent, physical understanding of the asymptotic behaviors. Then in Sec.~IV, we investigate the pattern of evolving to the center minimum. A special process called ``oscillating inflation'' is considered in detail. Finally, conclusions are given in Sec.~V. We adopt the following conventions: metric $\eta_{\mu\nu}=(-1,+1,+1,+1)$, natural unit $\hbar=c=1$ and $M_P\equiv1/\sqrt{8\pi G}=2.435\times10^{18}~\mathrm{GeV}=1$.

\section{Weyl scaling invariant $R^2+R^3$ model}

We start with the following Lagrangian for metric field $g_{\mu\nu}$, scalar field $\varphi$, and Weyl gauge field $W_\mu\equiv g_Ww_\mu$ with local scaling symmetry
\begin{equation}
	\frac{\mathcal L}{\sqrt{-g}}= \frac{1}{2}\left(\varphi^2\hat R+\alpha \hat R^2+\frac{\beta}{\varphi^2}\hat R^3\right)-\frac{\zeta}{2} D^\mu\varphi D_\mu\varphi-\frac{1}{4g_W^2}F_{\mu\nu}F^{\mu\nu}.
	\label{LWR3}
\end{equation}
Here $g$ is the determinant of metric, $\alpha, \beta$ and $\zeta$ are constant parameters, $D_\mu=\partial_\mu-W_\mu$ is the covariant derivative associated with scaling symmetry, $g_W$ is the coupling constant,  $F_{\mu\nu}\equiv\partial_\mu W_\nu-\partial_\nu W_\mu$ defines the invariant field strength of $W_\mu$, and $\hat R$ is the Ricci scalar defined by the local scaling invariant connection
\begin{equation}
	\begin{aligned}
	\hat{\Gamma}^\rho_{\mu\nu}=\frac{1}{2}g^{\rho\sigma}\left[(\partial_\mu+2W_\mu) g_{\sigma\nu}+(\partial_\nu+2W_\nu)g_{\mu\sigma}-(\partial_\sigma+2W_\sigma)g_{\mu\nu}\right].
	\end{aligned}
	\label{WGamma}
\end{equation}
Explicit calculation shows the relation between $\hat{R}$ and usual $R$ defined by metric field $g_{\mu\nu}$,
\begin{equation}
	\hat R=R-6W_\mu W^\mu-\frac{6}{\sqrt{-g}}\partial_\mu(\sqrt{-g}W^\mu).
	\label{WeylR}
\end{equation}
It is straightforward to verify the invariance of Eq.~(\ref{LWR3}) under the following Weyl scaling transformation
\begin{equation}
	\begin{aligned}
		\mathrm{metric}:~&g_{\mu\nu}\rightarrow g'_{\mu\nu}=f^2(x)g_{\mu\nu},\\
		\mathrm{scalar}:~&\phi\rightarrow \phi'=f^{-1}(x)\phi,\\
		\mathrm{Ricci~scalar}:~&\hat R\rightarrow \hat R'=f^{-2}(x)\hat R,\\
		\mathrm{Weyl~vector}:~&W_\mu\rightarrow W'_\mu=W_\mu-\partial_\mu \ln f(x),
		\label{CT}
	\end{aligned}
\end{equation}
where $f(x)$ is an arbitrary positive function. 

The purpose to explore the Lagrangian in Eq.~(\ref{LWR3}) is two-fold. Theoretically, such a $\hat{R}^3$ term constitutes as a simple extension of the $\hat{R}^2$ theory, motivated from perspective of effective field theories and also quantum loop corrections in more fundamental theories~\cite{Wu:2015wwa, Wu:2017urh, Wu:2021ign, Wu:2021ucc}. Phenomenologically, it is worthwhile to explore how such a term would modify the cosmological observations related to inflation, and evaluate the likelihood and robustness of the predictions in the lowest-order theories. 

\subsection{Formalism in Einstein frame}

General $f(R)$ gravity is equivalent to the Einstein gravity with a scalar field \cite{Whitt:1984pd,Barrow:1988xh}. In Ref.~\cite{Tang:2020ovf}, we have extended the proof in general scaling invariant $F(\hat{R},\varphi)$ gravity. We can explicitly show that by introducing an auxiliary scalar field $\chi$ and rewrite the high-order curvature terms as
\begin{equation}
	F(\hat R,\varphi)\equiv\varphi^2\hat R+\alpha \hat R^2+\frac{\beta}{\varphi^2}\hat R^3=F_{\hat{R}}(\hat{R}\rightarrow\chi^2,\varphi)(\hat R-\chi^2)+F(\hat{R}\rightarrow\chi^2,\varphi).
	\label{FR}
\end{equation}
Here $F_{\hat{R}}$ denotes the derivative over $\hat{R}$, $F_{\hat{R}}=\partial F(\hat R,\varphi) /\partial \hat{R}$. We can verify that the equivalence relation $\chi^2=\hat R$ can be obtained from the Euler-Lagrange equation, $\frac{\delta\mathcal L}{\delta\chi}=0$. Substituting Eq.~(\ref{FR}) into Eq.~(\ref{LWR3}), we find
\begin{equation}
	\begin{aligned}
		\frac{\mathcal L}{\sqrt{-g}}=\frac{1}{2}\left(\varphi^2+2\alpha\chi^2+\frac{3\beta}{\varphi^2}\chi^4\right)\hat R-\frac{1}{2}\left(\alpha \chi^4+\frac{2\beta}{\varphi^2}\chi^6\right)-\frac{\zeta}{2}D^\mu\varphi D_\mu\varphi-\frac{1}{4g_W^2}F_{\mu\nu}F^{\mu\nu}.
		\label{L2}
	\end{aligned}
\end{equation}
Now we have demonstrated that linearization of $\hat{R}$ has led to the nonminimal coupling of the scalar field, $\chi$. 

We can transform the above Lagrangian into the Einstein frame by making a Weyl or conformal transformation of the metric field. However, we note that scaling invariance is still preserved in our model with $\chi\rightarrow \chi'=f^{-1}(x)\chi$. Therefore, we can directly normalize the coefficient before the Ricci scalar as 
\begin{equation}\label{eq:gaugefix}
	\varphi^2+2\alpha\chi^2+ 3\beta\chi^4/\varphi^2=1,
\end{equation}
due to the scaling invariance of Eq.~(\ref{L2}). This is equivalent to making a Weyl transformation with $f(x)=\sqrt{\varphi^2+2\alpha\chi^2+3\beta\chi^4/\varphi^2}$ in Eq.~(\ref{CT}). Further dropping the total derivative term in Eq.~(\ref{WeylR}) due to its null surface integral, we can write the Lagrangian as
\begin{equation}
	\begin{aligned}
		\frac{\mathcal L}{\sqrt{-g}}=&\frac{1}{2}R-\frac{\zeta}{2}D^\mu\varphi D_\mu\varphi-V(\varphi)-\frac{1}{4g_W^2}F_{\mu\nu}F^{\mu\nu}-3W^\mu W_\mu \\
		=&\frac{R}{2}-\frac{\partial^\mu\varphi\partial_\mu\varphi}{2/\zeta+\varphi^2/3}-V(\varphi)-\frac{1}{4g_W^2}F_{\mu\nu}F^{\mu\nu}-\frac{6+\zeta\varphi^2}{2}\left[W_\mu-\frac{\partial_\mu\ln|6+\zeta\varphi^2|}{2}\right]^2,
		\label{L3}
	\end{aligned}
\end{equation}
with the scalar potential
\begin{equation}
	\begin{aligned}
		V(\varphi) =\frac{\alpha}{2}\chi^4+\frac{\beta}{\varphi^2} \chi^6 =\frac{\alpha}{6\beta}\left(\varphi^4-\varphi^2\right)+\frac{\alpha^3\varphi^4}{27\beta^2}\left[\left(1-\frac{3\beta}{\alpha^2}\left(1-\varphi^{-2}\right)\right)^{3/2}-1\right],
		\label{VWR3}
	\end{aligned}
\end{equation}
where we have solved $\chi$ from Eq.~(\ref{eq:gaugefix})
\begin{equation}
	\chi^2=\frac{\alpha\varphi^2}{3\beta}\left[\sqrt{1-\frac{3\beta}{\alpha^2}\left(1-\varphi^{-2}\right)}-1\right].
	\label{root2}
\end{equation}

It is now clear that we have a minimally coupled scalar $\varphi$ with a noncanonical kinetic term. To further simplifying the theoretical formalism, we introduce the following redefinitions for the scalar and the Weyl gauge field
\begin{equation}
	\varphi^2\equiv
	\begin{cases}
		\frac{6}{|\zeta|}\sinh^2\left(\frac{\pm\Phi}{\sqrt 6}\right)~\mathrm{for}~ \zeta>0,\\
		\frac{6}{|\zeta|}\cosh^2\left(\frac{\pm\Phi}{\sqrt 6}\right)~\mathrm{for}~ \zeta<0,
	\end{cases}
	\label{Phi}
\end{equation}
\begin{equation}
	\tilde W_\mu\equiv W_\mu-\frac{1}{2}\partial_\mu\ln|6+\zeta\varphi^2|\equiv g_W\tilde w_\mu.
\end{equation}
Then the final Lagrangian turns into a more compact form
\begin{equation}
	\frac{\mathcal L}{\sqrt{-g}}=\frac{1}{2}R-\frac{1}{2}\partial^\mu\Phi\partial_\mu\Phi-V(\Phi)-\frac{1}{4g_W^2}\tilde F_{\mu\nu}\tilde F^{\mu\nu}-\frac{1}{2}m^2(\Phi)\tilde W^\mu\tilde W_\mu,
	\label{L4}
\end{equation}
with the mass term of Weyl gauge field
\begin{equation}
	m^2(\Phi)=
	\begin{cases}
		+6\cosh^2\left(\frac{\Phi}{\sqrt{6}}\right) ~\mathrm{for}~ \zeta>0,\\
		-6\sinh^2\left(\frac{\Phi}{\sqrt{6}}\right) ~\mathrm{for}~ \zeta<0.\\
	\end{cases}
	\label{m2}
\end{equation}
We should note that $m^2(\Phi)$ is negative when $\zeta<0$. Therefore, to avoid the Weyl gauge boson becoming tachyonic in this case, it requires some other mechanisms to obtain a real mass, for example, introducing other scalar field, which we do not explore in this paper. For viable inflation, both positive and negative are possible, as we shall show later.

In the above discussion, we have demonstrated that Weyl scaling invariant $\hat{R}^2+\hat{R}^3$ model can be written equivalently as the Einstein gravity coupled with a self-interacting scalar $\Phi$ and a massive vector $\tilde W_\mu$ with a field-dependent mass. This conclusion is also true for any Weyl scaling invariant model of gravity with high-order curvature $\hat R^n$ as the above formalism applies straightforwardly. It is also worthwhile to point out that Weyl vector boson can serve as a dark matter candidate~\cite{Tang:2020ovf, Tang:2019uex, Tang:2019olx}, with details of the relic abundance being discussed in~\cite{Wang:2022ojc}. In this paper, we shall concentrate on the scalar potential Eq.~(\ref{VWR3}) and discuss the viable inflation scenarios with the presence of $\hat{R}^3$.

\subsection{Effective scalar potentials}

\begin{figure}
	\centering
	\includegraphics[width=0.95\textwidth]{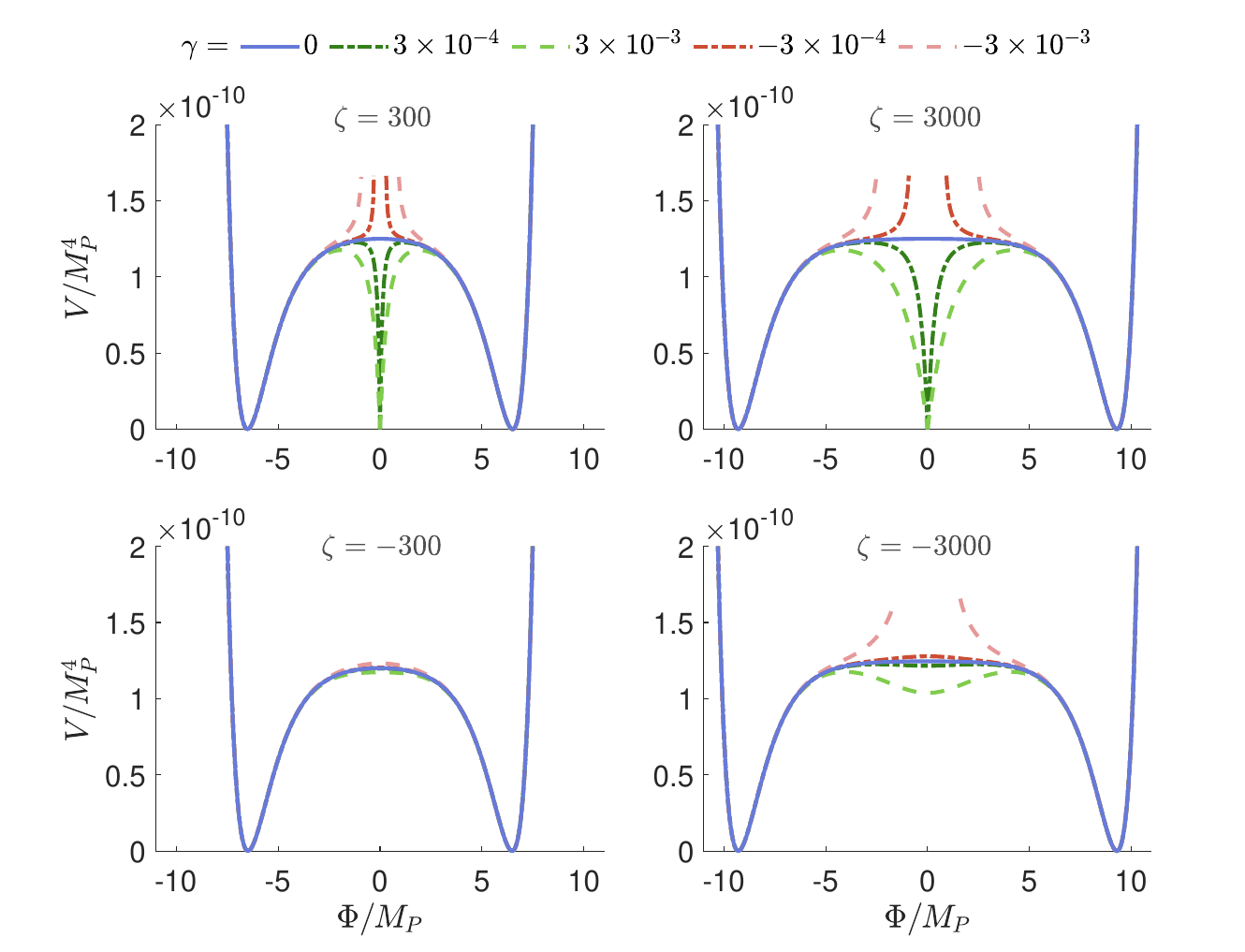}
	\caption{Effective potentials of Weyl $R^2+R^3$ model with $\alpha=10^9$ and various $\gamma$ and $\zeta$. Here we only depict the real ranges of potentials.}
	\label{VR3W}
\end{figure}

There are two necessary requirements for the potential Eq.~(\ref{VWR3}). The first one is $\varphi^2>0$ since $\varphi$ is a real scalar field. The other is $1-\frac{3\beta}{\alpha^2}\left(1-\frac{1}{\varphi^2}\right)\geq0$, otherwise an imaginary potential will emerge. Consequently, there are some constraints on the parameters and the viable value of $\Phi$. We can rewrite the second requirement as
\begin{equation}
	\begin{aligned}
		&\sinh^2\left(\frac{\pm\Phi}{\sqrt{6}}\right)\geq~\mathrm{or}~\leq\frac{|\zeta|}{6-2\alpha^2/\beta},~\mathrm{for}~\zeta>0,\\
		&\cosh^2\left(\frac{\pm\Phi}{\sqrt{6}}\right)\geq~\mathrm{or}~\leq\frac{|\zeta|}{6-2\alpha^2/\beta},~\mathrm{for}~\zeta<0,
	\end{aligned}
\end{equation}
where `` $\geq$ '' for $\beta<\frac{\alpha^2}{3}$ and `` $\leq$ '' for $\beta\geq\frac{\alpha^2}{3}$. For convenience, we define $\lambda\equiv\sqrt{\frac{|\zeta|}{6-2\alpha^2/\beta}}$ and $\gamma\equiv\frac{3\beta}{\alpha^2}$ (only for $\beta<0$ or $\beta>\alpha^2/3$ cases), then discuss the possible ranges of the potential corresponding to different parameters. The results are listed in the Table~{\ref{Tab}}. To ensure the theoretical stability, we require that $\Phi$ can only evolve within these ranges where the potential is real. Figure~\ref{VR3W} shows some instances of the scalar potential for several values of $\zeta$ and $\gamma$.

\begin{table}[t]
	\caption{Effective potential range of the Weyl $R^2+R^3$ model.}
	\begin{tabular}{|p{3cm}<{\centering}|p{5cm}<{\centering}|p{5cm}<{\centering}|}\hline
		$\zeta$ & $\gamma$ or $\beta$ & real $V(\varphi)$ \\ \hline
		
		\multirow{3}{*}{$\zeta>0$} & $\gamma\geq1$ & $|\Phi|\leq\sqrt{6}\sinh^{-1}\lambda$ \\ \cline{2-3}
		\multirow{3}{*}{} & $0\leq\gamma<1$ & fully real \\ \cline{2-3}
		\multirow{3}{*}{} & $\gamma<0$ & $|\Phi|\geq\sqrt{6}\sinh^{-1}\lambda$ \\ \hline
		
		\multirow{3}{*}{$-6<\zeta<0$} & $\gamma>\frac{1}{1+\zeta/6}$ & fully imaginary \\ \cline{2-3}
		\multirow{3}{*}{} &$1<\gamma\leq\frac{1}{1+\zeta/6}$ & $|\Phi|\leq\sqrt{6}|\cosh^{-1}\lambda|$ \\ \cline{2-3}
		\multirow{3}{*}{} & $\gamma\leq1$ & fully real \\ \hline
		
		\multirow{3}{*}{$\zeta\leq-6$} & $\gamma\geq1$ & $|\Phi|\leq\sqrt{6}|\cosh^{-1}\lambda|$ \\ \cline{2-3}
		\multirow{3}{*}{} & $\frac{1}{1+\zeta/6}<\gamma<1$ & fully real \\ \cline{2-3}
		\multirow{3}{*}{} & $\gamma\leq\frac{1}{1+\zeta/6}$ & $|\Phi|\geq\sqrt{6}|\cosh^{-1}\lambda|$ \\ \hline
	\end{tabular}
	\label{Tab}
\end{table}

We first discuss the case of positive $\zeta$. When $\gamma=0$, it is a hilltoplike potential with two minima at $\Phi=\pm\sqrt{6}\sinh^{-1}\sqrt{\frac{\zeta}{6}}$. However, as long as there is a tiny cubic curvature, whether positive or negative, the shape of potential will be affected significantly. When $\gamma>0$, the potential turns to decrease near $\Phi=0$, and a third vacuum can form there. This behavior is transparent, because when $\zeta>0$, $\Phi=0$ corresponds to $\varphi^2=0$ according to Eq.~(\ref{Phi}), then substituting it in Eq.~(\ref{VWR3}) will obtain $V|_{\Phi=0}=0$. When $\gamma<0$, the potential turns to rise near $\Phi=0$ and become imaginary and unphysical in $-\sqrt{6}\sinh^{-1}\lambda<\Phi<\sqrt{6}\sinh^{-1}\lambda$, which has been listed in Table.~\ref{Tab}.

Next, we switch to the case of negative $\zeta$. It is evident in Fig.~\ref{VR3W} that when $\zeta<0$ and $|\zeta|$ or $|\gamma|$ is relatively small, the modification of $\hat R^3$ term on the Weyl $R^2$ potential is moderate, unlike the dramatic change near $\Phi=0$ in the case of positive $\zeta$. This is because the mapping of $\Phi\Rightarrow\varphi^2$ does not cover the interval of $\varphi^2<1$ for $\zeta<0$ according to Eq.~(\ref{Phi}). In other words, for negative $\zeta$ with modest $|\gamma|$, $\Phi\rightarrow 0$ does not lead to $\varphi^2\rightarrow 0$, which brings the violent behavior of the potential around here in the case of $\zeta>0$. However, when $\zeta$ is excessively negative or $|\gamma|$ is large enough, the violent variation will reappear to a certain extent. For $\gamma > 0$, the potential will return to a downward trend near $\Phi=0$, albeit there is no true vacuum formed (but a false vacuum is formed). And for excessively negative $\gamma$, the imaginary potential will reappear in the range of $-\sqrt{6}|\cosh^{-1}\lambda|<\Phi<\sqrt{6}|\cosh^{-1}\lambda|$, which we have listed this situation in Table.~\ref{Tab} (see $\zeta\leq-6$ with $\gamma\leq\frac{1}{1+\zeta/6}$ case).

Generally, inflation takes place when the potential is flat and $\Phi$ evolves to the vacuum ($\Phi|_{V=0}$). The cosmological observations would restrict the potential and the initial value $\Phi_i$ when inflation starts, here the $\Phi_i$ is defined as the value when the comoving horizon of the inflationary universe shrinks to the same size as today. 

For $\zeta>0$ and $\gamma>0$, the scalar potential contains three separate vacua, one lying at the center and the other two at both sides. Therefore, there are two different viable inflationary patterns. One pattern refers to the evolution into the central minimum, and the other into the side minima. We can calculate the value of $\Phi$ which corresponds to the hilltop of the potential in this case
\begin{equation}
	\Phi_h=\pm\sqrt{6}\sinh^{-1}\sqrt{\frac{\zeta}{12}\frac{\sqrt{3\gamma}-2\gamma}{3-4\gamma}},~\zeta>0,~\gamma>0
	\label{Phitop+}
\end{equation}
which is the critical point of two inflationary patterns. Neglecting the velocity, if the initial value of inflation field satisfies $|\Phi_i|>|\Phi_h|$, it will evolve toward the side vacua. If $|\Phi_i|<|\Phi_h|$ at the beginning, the inflation field will evolve toward the central minimum. There is another point worth noting. The potential at $\Phi=0$ in this case has no continuous left and right derivatives. This seems to be problematic when the inflaton falls into the central minimum. However, if we consider the existence of higher-order curvature, e.g., $\hat R^4/\varphi^4$, there will be a rounded bottom at $\Phi=0$, and if the higher-order curvature is small, its influence will only concentrate around $\Phi=0$ without affecting the physical quantities of slow-roll inflation (see Appendix \ref{ap1} in detail).

For other cases of $\zeta$ and $\gamma$, there are only the global side minima. Hence the only feasible inflationary pattern is that $\Phi$ evolves to either one of the side minimum. The initial value $\Phi_i$ has to correspond to a real potential, and when there is a false vacuum in $\zeta<0$ case, it requires a large enough $|\Phi_i|$ outside two local maxima of the potential to ensure the gradient of $V(\Phi_i)$ toward the true vacuum. Next, we are going to discuss the inflation in these two patterns respectively.

\section{Inflation to the side}

In this inflation pattern, $\varphi^2$ [defined as Eq.~(\ref{Phi})] is usually not very close to 0, and as we shall show later, observations generally would require an extremely small cubic curvature, namely $|\gamma|\ll 1$. Therefore in many cases, $\left|\gamma(1- \varphi^{-2})\right|\ll1$ is satisfied. Under this condition, we are able to have analytical treatment and expand the potential Eq.~(\ref{VWR3}) as
\begin{align}\label{Vexpand}
		V(\varphi)=&\frac{\varphi^4-\varphi^2}{2\alpha\gamma}+\frac{\varphi^4}{3\alpha\gamma^2}\left[-\frac{3\gamma}{2}\left(1-\frac{1}{\varphi^2}\right)+\frac{3\gamma^2}{8}\left(1-\frac{1}{\varphi^2}\right)^2+\frac{\gamma^3}{16}\left(1-\frac{1}{\varphi^2}\right)^3+\mathcal O\left(\frac{\gamma^4}{\varphi^8}\right)\right]\nonumber\\
		=&\frac{1}{8\alpha}\left(1-\varphi^2\right)^2\left[1+\frac{\gamma}{6}\left(1-\frac{1}{\varphi^2}\right)+\mathcal O\left(\frac{\gamma^2}{\varphi^4}\right)\right].
\end{align}
Then with Eq.~(\ref{Phi}), we derive
\begin{align}\label{V}
		V(\Phi)=
		\begin{cases}
		\frac{1}{8\alpha}\left[1-\frac{6}{|\zeta|}\sinh^2\left(\frac{\Phi}{\sqrt{6}}\right)\right]^2\left[1+\frac{\gamma}{6}\left(1-\frac{|\zeta|}{6}\mathrm{csch}^2\left(\frac{\Phi}{\sqrt{6}}\right)\right)+\mathcal O(\gamma^2)\right] ~\mathrm{for}~ \zeta>0,\\
		\frac{1}{8\alpha}\left[1-\frac{6}{|\zeta|}\cosh^2\left(\frac{\Phi}{\sqrt{6}}\right)\right]^2\left[1+\frac{\gamma}{6}\left(1-\frac{|\zeta|}{6}\mathrm{sech}^2\left(\frac{\Phi}{\sqrt{6}}\right)\right)+\mathcal O(\gamma^2)\right] ~\mathrm{for}~ \zeta<0.
		\end{cases}
\end{align}
The first term is exactly the effective potential of Weyl $\hat{R}^2$, which has been shown in \cite{Tang:2020ovf,Wang:2022ojc}, and the rest originates from the cubic curvature term $\hat{R}^3$, to the leading order of $\gamma$. Next we shall calculate the inflationary physical quantities, the spectral index $n_s$ and tensor-to-scalar ratio $r$, and contrast them with the latest observations. We first give an analytical calculation for two limiting cases, then show the full numerical results for general cases.

\subsection{Analytical approach of $\gamma\rightarrow0$ case}

We first discuss the $\gamma\rightarrow0$ case and show how $\zeta$ affects  $n_s$ and $r$. The slow-roll parameters in this case can be derived as
\begin{equation}
	\begin{aligned}
		\epsilon\equiv\frac{1}{2}\left[\frac{V'(\Phi)}{V}\right]^2=\frac{12\sinh^2\left(\frac{2\Phi}{\sqrt6}\right)}{\left[|\zeta+3|-3-6\sinh^2\left(\frac{\Phi}{\sqrt6}\right)\right]^2},
		\label{eps}
	\end{aligned}
\end{equation}
\begin{equation}
	\begin{aligned}
		\eta\equiv\frac{V''(\Phi)}{V}=\frac{12\cosh\left(\frac{4\Phi}{\sqrt6}\right)-4|\zeta+3|\cosh\left(\frac{2\Phi}{\sqrt6}\right)}{\left[|\zeta+3|-3-6\sinh^2\left(\frac{\Phi}{\sqrt6}\right)\right]^2}.
		\label{eta}
	\end{aligned}
\end{equation}
Generally, the slow-roll inflation occurs when $\epsilon$ and $|\eta|$ is small enough, and it will end when any of them evolves to $\sim 1$. For the situation we are concerned with, $\epsilon$ breaks the slow-roll limit before the other. Thus we derive the value of $\Phi$ when inflation ends according to $\epsilon=1$
\begin{align}
		\Phi_e=\sqrt{\frac{3}{2}}\ln\left(\frac{2\sqrt{|\zeta+3|^2+3}}{\sqrt{3}}-|\zeta+3|+\sqrt{\frac{7}{3}|\zeta+3|^2-\frac{4|\zeta+3|}{\sqrt{3}}\sqrt{|\zeta+3|^2+3}+3}\right).
\end{align}
When $|\zeta|>\mathcal O(10^2)$, which is a preferred range by the observational constraints as we will show shortly, the above equation can be approximated as
\begin{align}\label{Phie}
	\Phi_e\simeq\sqrt{\frac{3}{2}}\ln\left[\frac{1}{\sqrt3}\left(2+\sqrt{7-4\sqrt3}-\sqrt3\right)|\zeta+3|\right]
		\simeq\sqrt{\frac{3}{2}}\ln\left(0.3094|\zeta+3|\right).
\end{align}
It is now clear that when $|\zeta|$ is large enough, $\Phi_e$ will be almost independent of the sign of $\zeta$.

Next, we calculate initial value $\Phi_i$, which is defined when the size of comoving horizon during inflation shrinks to the present size. We first focus on the $e$-folding number of the slow-roll inflation
\begin{equation}
	N\equiv\ln\frac{a_e}{a_i}\simeq\int^{\Phi_e}_{\Phi_i}\frac{\mathrm d\Phi}{\sqrt{2\epsilon}},
	\label{N}
\end{equation}
where $a_{i/e}\equiv a(\Phi_{i/e})$ is the cosmic scale factor when inflation starts/ends. Substituting Eq.~(\ref{eps}) into it, we find
\begin{equation}
	\begin{aligned}
		N&=\frac{(|\zeta+3|-3)\ln\left[\tanh\left(\frac{\Phi}{\sqrt6}\right)\right]-6\ln\left[\cosh\left(\frac{\Phi}{\sqrt6}\right)\right]}{4}\Bigg|^{\Phi_e}_{\Phi_i}\\
		&=\frac{|\zeta+3|-3}{4}\ln\left[\frac{\tanh\left[\frac{1}{2}\ln(0.3094|\zeta+3|)\right]}{\tanh\left(\frac{\Phi_i}{\sqrt6}\right)}\right]-\frac{3}{2}\ln\left[\frac{\cosh\left[\frac{1}{2}\ln(0.3094|\zeta+3|)\right]}{\cosh\left(\frac{\Phi_i}{\sqrt6}\right)}\right].
	\end{aligned}
	\label{NN}
\end{equation}
For the circumstances we are concerned with, namely $N\sim(50,60)$ and $|\zeta|>\mathcal O(10^2)$, the second term of Eq.~(\ref{NN}) is much smaller than the first term, and it can be estimated as $\sim-2.3$. Thus we derive
\begin{equation}
	\Phi_i\simeq\sqrt6\tanh^{-1}\left[\left(1-\frac{2}{0.3094|\zeta+3|+1}\right)e^{\frac{-4(N+2.3)}{|\zeta+3|-3}}\right]\equiv\sqrt6\tanh^{-1}\Omega(\zeta,N).
	\label{Phii}
\end{equation}
Here we have defined $\Omega(\zeta,N)$ for later convenience.

When $|\zeta|\gg 4N$, it can be further approximated as $\Phi_i\simeq\sqrt{\frac{3}{2}}\ln\frac{|\zeta|}{2N+7.8}$. Substituting Eq.~(\ref{Phii}) into Eq.~(\ref{eps}) and (\ref{eta}), we find
\begin{equation}
	\begin{aligned}
		\epsilon_i=\frac{48\Omega^2}{\left[(\Omega^2-1)|\zeta+3|+3(\Omega^2+1)\right]^2},
	\end{aligned}
\end{equation}
\begin{equation}
	\begin{aligned}
		\eta_i=&\frac{4\left[(\Omega^4-1)|\zeta+3|+3(\Omega^4+6\Omega^2+1)\right]}{\left[(\Omega^2-1)|\zeta+3|+3(\Omega^2+1)\right]^2}.
	\end{aligned}
\end{equation}

\begin{figure}[t]
	\centering
	\includegraphics[width=0.8\textwidth]{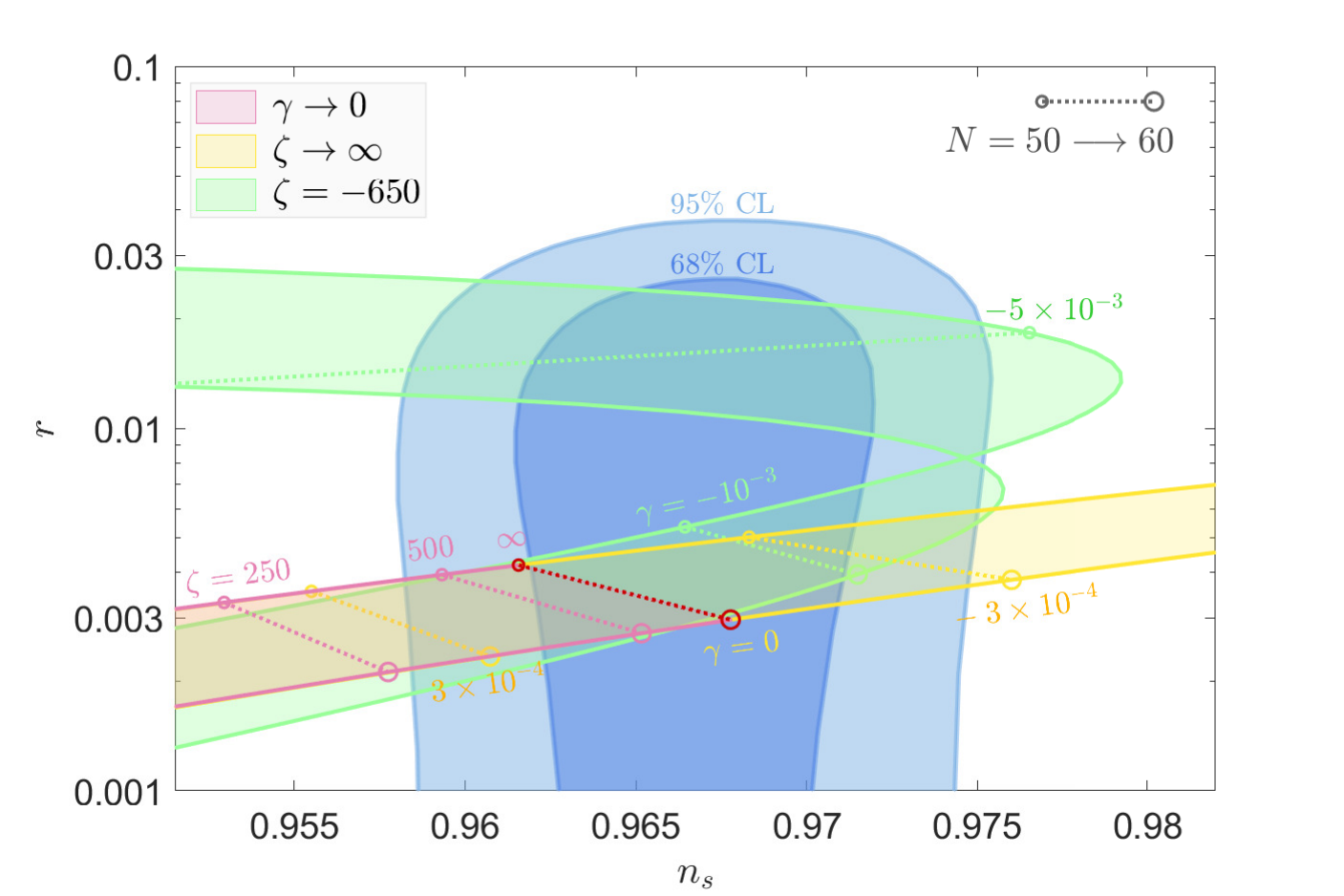}
	\caption{The predictions of spectral index $n_s$ combined with tensor-to-scalar ratio $r$ in the Weyl $R^2+R^3$ model with $e$-folding number $N\sim(50,60)$. The pink area shows the results in the $\gamma\rightarrow0$ case with various $\zeta$. The yellow and green areas respectively show the $\zeta\rightarrow\infty$ and $\zeta=-650$ cases with various $\gamma$. The red line is the result with both $\gamma\rightarrow0$ and $\zeta\rightarrow\infty$, which is equivalent to the Starobinsky model. The blue area is the latest observation constraint given by the BICEP/$Keck$ collaboration \cite{BICEP:2021xfz}.}
	\label{r_ns}
\end{figure}

As a result, the tensor-to-scalar ratio $r$ and spectral index $n_s$ of inflationary perturbations in the $\gamma\rightarrow0$ limit are finally calculated as
\begin{equation}
	r=16\epsilon_i=\frac{768\Omega^2}{\left[(\Omega^2-1)|\zeta+3|+3(\Omega^2+1)\right]^2},
\end{equation}
\begin{equation}
	\begin{aligned}
		n_s=1-6\epsilon_i+2\eta_i=1+\frac{8(\Omega^4-1)|\zeta+3|+24(\Omega^4-6\Omega^2+1)}{\left[(\Omega^2-1)|\zeta+3|+3(\Omega^2+1)\right]^2}.
	\end{aligned}
\end{equation}
For $N\sim(50,60)$ and $|\zeta|>\mathcal O(10^2)$, We can approximate the expressions as
\begin{equation}
	r\simeq r^*-\frac{54}{\zeta^2},
	\label{r1}
\end{equation}
\begin{equation}
	n_s\simeq n_s^*-\frac{11N}{\zeta^2},
	\label{ns1}
\end{equation}
where
\begin{equation}
	r^*\simeq\dfrac{12}{(N+3.55)^2},\; n_s^*\simeq1-\dfrac{2}{N+3.55}-\dfrac{3}{(N+3.55)^2}
\end{equation}
are the predictions of Starobinsky model (see Appendix \ref{ap2} for an analytical derivation.). Thus it is evident that the predictions of inflationary perturbations in our model will converge to that of Starobinsky model when $\gamma\rightarrow 0$ and $\zeta\rightarrow \infty$. As $|\zeta|$ decreases, the value of $r$ and $n_s$ will also decrease. We show this trend as the pink area in Fig.~\ref{r_ns}. According to the latest observation \cite{BICEP:2021xfz}, the lower limit of $n_s$ has been constrained to $\sim 0.959$, hence it requires $|\zeta|>270$ in this $\gamma\rightarrow 0$ case.

\subsection{Analytical approach of $\zeta\rightarrow\infty$ case}

Now we discuss the $\zeta\rightarrow\infty$ case and show how $\gamma$ affects $r$ and $n_s$. When $\zeta$ is large enough, the potential is greatly widened. The side vacua are far away from 0 and so are $\Phi_i$ and $\Phi_e$ (e.g., $\Phi_i\sim 5.4M_P$, $\Phi_e\sim 9.8M_P$ for $\zeta=10^4$). Therefore Eq.~(\ref{Phi}) can be approximated as
\begin{equation}
	\begin{aligned}
		\varphi^2=\frac{6}{|\zeta|}\left(\frac{e^{\Phi/\sqrt{6}}\pm e^{-\Phi/\sqrt{6}}}{2}\right)^2\simeq e^{{\sqrt{2/3}\left[\Phi-\sqrt{3/2}\ln(2|\zeta|/3)\right]}}\equiv e^{\sqrt{2/3}(\Phi-\Phi_0)}.
	\end{aligned}
	\label{Phiappro}
\end{equation}
Here and after, without losing generality, we may choose to evolve in the positive $\Phi$ region, and denote $\Phi_0$ as the minimum in this region. Substituting it into Eq.~(\ref{Vexpand}), we have the scalar potential for $\Phi\gg0$
\begin{equation}
	\begin{aligned}
		V(\Phi)=\frac{1}{8\alpha}\left(1-e^{\sqrt{2/3}(\Phi-\Phi_0)}\right)^2\left[1+\frac{\gamma}{6}\left(1-e^{-\sqrt{2/3}(\Phi-\Phi_0)}\right)+\mathcal O(\gamma^2)\right].
	\end{aligned}
	\label{V3}
\end{equation}
Ignoring the $\mathcal O(\gamma^2)$ terms, we give an approximate expression for the slow-roll parameters
\begin{equation}
	\begin{aligned}
		\epsilon\equiv\frac{1}{2}\left[\frac{V'(\Phi)}{V}\right]^2\simeq\frac{\left[\gamma e^{\sqrt{2/3}(\Phi-\Phi_0)}-2(\gamma+6)e^{\sqrt{8/3}(\Phi-\Phi_0)}+\gamma\right]^2}{3\left[e^{\sqrt{2/3}(\Phi-\Phi_0)}-1\right]^2\left[\gamma-(\gamma+6)e^{\sqrt{2/3}(\Phi-\Phi_0)}\right]^2},
		\label{eps2}
	\end{aligned}
\end{equation}
\begin{equation}
	\begin{aligned}
		\eta\equiv\frac{V''(\Phi)}{V}\simeq\frac{6(\gamma+4)e^{\sqrt{8/3}(\Phi-\Phi_0)}-8(\gamma+6)e^{\sqrt{6}(\Phi-\Phi_0)}+2\gamma}{3\left[e^{\sqrt{2/3}(\Phi-\Phi_0)}-1\right]^2\left[\gamma-(\gamma+6)e^{\sqrt{2/3}(\Phi-\Phi_0)}\right]}.
		\label{eta2}
	\end{aligned}
\end{equation}
In this case, the slow-roll inflation also ends at $\epsilon\sim1$. To find the expression of $\Phi_e$, we further approximate Eq.~(\ref{eps2}) as
\begin{equation}
	\begin{aligned}
		\epsilon\simeq\frac{e^{-\sqrt{8/3}(\Phi-\Phi_0)}\left(\gamma-12e^{\sqrt{8/3}(\Phi-\Phi_0)}\right)^2}{108\left(e^{\sqrt{2/3}(\Phi-\Phi_0)}-1\right)^2}.
	\end{aligned}
	\label{eps2s}
\end{equation}
Then $\Phi_e$ can be derived as
\begin{equation}
	\begin{aligned}
		\Phi_e=\Phi_0-\sqrt{\frac{3}{2}}\ln\left[\frac{\sqrt3}{\gamma}\left(\sqrt{2(2+\sqrt3)\gamma+9}-3\right)\right].
	\end{aligned}
\end{equation}
If $\gamma$ is extremely small, we will find $\Phi_e\simeq\Phi_0-0.94M_P$.

Next, we derive the analytic formula for $\Phi_i$ in this case. The $e$-folding number of the slow-roll inflation can be calculated with Eq.~(\ref{eps2s}) as
\begin{equation}
	\begin{aligned}
		N=-\sqrt\frac{27}{4\gamma}\tanh^{-1}\left[\sqrt\frac{\gamma}{12}e^{-\sqrt{\frac{2}{3}}(\Phi-\Phi_0)}\right]-\frac{3}{8}\ln\left[12-\gamma e^{-\sqrt{\frac{8}{3}}(\Phi-\Phi_0)}\right]-\frac{\sqrt6}{4}(\Phi-\Phi_0)\bigg|^{\Phi_e}_{\Phi_i}.
	\end{aligned}
	\label{N2}
\end{equation}
Considering $N\sim(50,60)$ and $\gamma<\mathcal O(10^{-3})$, the first term of the integral is dominant, while the rest are the marginal terms which can be approximately  treated as a constant, $\sim-2.7$. Hence we have
\begin{equation}
	\begin{aligned}
		N\simeq\sqrt\frac{27}{4\gamma}\left[\tanh^{-1}\left(\sqrt\frac{\gamma}{12}e^{-\sqrt{2/3}(\Phi_i-\Phi_0)}\right)-\tanh^{-1}\left(\sqrt\frac{\gamma}{12}e^{-\sqrt{2/3}(\Phi_e-\Phi_0)}\right)\right]-2.7,
	\end{aligned}
\end{equation}
and derive
\begin{equation}
	\begin{aligned}
		\Phi_i&=\Phi_0-\sqrt{\frac{3}{2}}\ln\left|\sqrt\frac{12}{\gamma}\tanh\left[\tanh^{-1}\left(\sqrt\frac{\gamma}{12}e^{-\sqrt{2/3}(\Phi_e-\Phi_0)}\right)+\sqrt\frac{4\gamma}{27}(N+2.7)\right]\right|\\
		&\simeq\Phi_0-\sqrt{\frac{3}{2}}\ln\left|\sqrt\frac{12}{\gamma}\tanh\left[\tanh^{-1}\left(0.622\sqrt\gamma\right)+\sqrt\frac{4\gamma}{27}(N+2.7)\right]\right|\\
		&\equiv\Phi_0-\sqrt{\frac{3}{2}}\ln\Theta(\gamma,N),
	\end{aligned}
\end{equation}
where we have defined $\Theta(\gamma,N)$ for later convenience. Then substituting it into Eq.~(\ref{eps2}) and (\ref{eta2}), we find
\begin{equation}
	\begin{aligned}
		\epsilon_i=\frac{\left[\gamma\Theta(1+\Theta)-2(\gamma+6)\right]^2}{3\left[1-\Theta\right]^2\left[\gamma\Theta-(\gamma+6)\right]^2},
	\end{aligned}
\end{equation}
\begin{equation}
	\begin{aligned}
		\eta_i=\frac{2\gamma\Theta^3+6(\gamma+4)\Theta-8(\gamma+6)}{3\left[1-\Theta\right]^2\left[\gamma\Theta-(\gamma+6)\right]}.
	\end{aligned}
\end{equation}

Finally, we derive $r$ and $n_s$ of the inflationary perturbations in the $\zeta\rightarrow\infty$ limit
\begin{equation}
	r=16\epsilon_i=\frac{16\left[\gamma\Theta(1+\Theta)-2(\gamma+6)\right]^2}{3\left[1-\Theta\right]^2\left[\gamma\Theta-(\gamma+6)\right]^2},
\end{equation}
\begin{equation}
	\begin{aligned}
		n_s=1-6\epsilon_i+2\eta_i=1-\frac{2\left[\gamma\Theta(1+\Theta)-2(\gamma+6)\right]^2}{\left[1-\Theta\right]^2\left[\gamma\Theta-(\gamma+6)\right]^2}+4\frac{\gamma\Theta^3+3(\gamma+4)\Theta-4(\gamma+6)}{3\left[1-\Theta\right]^2\left[\gamma\Theta-(\gamma+6)\right]}.
	\end{aligned}
\end{equation}
If $\gamma$ is extremely small, smaller than $\mathcal O(10^{-4})$, the above expressions can be linearly approximated as
\begin{equation}
	r\simeq r^*-2.4\gamma,
\end{equation}
\begin{equation}
	n_s\simeq n_s^*-0.42\gamma N,
\end{equation}
where $r^*$ and $n_s^*$ have been defined in the last paragraph of Sec.~III.A. We can see that compared with the predictions of Starobinsky model, a positive $\gamma$ will reduce both $r$ and $n_s$, while a negative $\gamma$ will increase them. We show this trend as the yellow area in Fig.~\ref{r_ns}. It is manifest that the observations have constrained $|\gamma|\lesssim 5\times 10^{-4}$ in this $\zeta\rightarrow\infty$ case. Actually, this result agrees with other numerical investigations of the $R^3$-extended Starobinsky model~\cite{Huang:2013hsb, Asaka:2015vza, Pi:2017gih, Cheong:2020rao, Rodrigues-da-Silva:2021jab, Ivanov:2021chn, Shtanov:2022pdx, Modak:2022gol}, since the potential Eq.~(\ref{V3}) is the same as the $R^3$-extended Starobinsky model with a vacuum shift. Moreover, compared with Eq.~(\ref{r1}) and (\ref{ns1}), we note that the predictions of $r$ and $n_s$ in the $\gamma\rightarrow 0$ case is similar to that of the $\zeta\rightarrow\infty$ and $\gamma>0$ case with a simple replacement of $\gamma\leftrightarrow\frac{24}{\zeta^2}$. This can be seen more clearly from Fig.~\ref{r_ns}, where the pink area overlaps with the yellow area with $\gamma>0$. 

\subsection{General cases}

\begin{figure}
	\centering
	\includegraphics[width=0.85\textwidth]{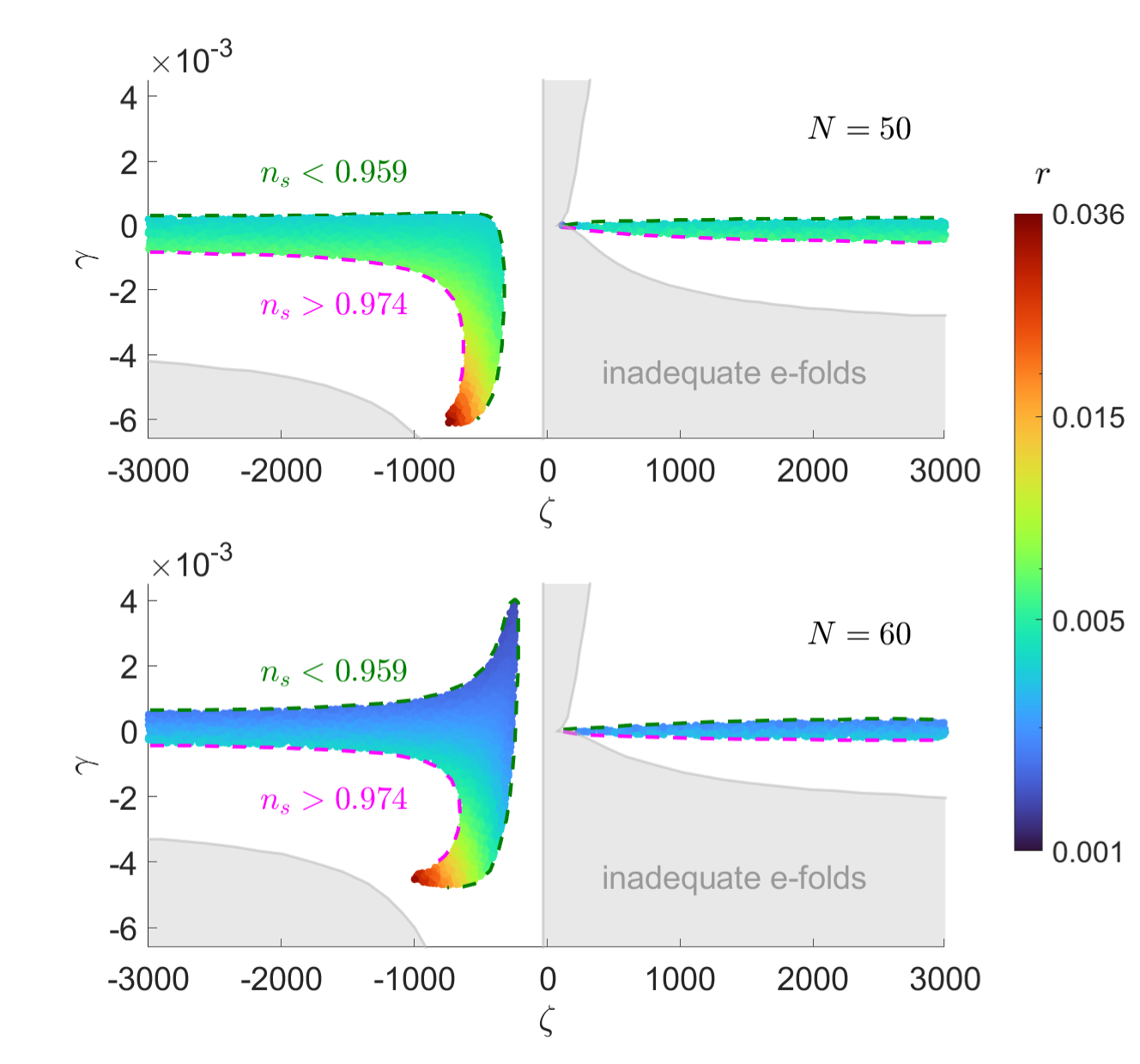}
	\caption{Possible parameter space for Weyl $R^2+R^3$ model when $\Phi$ evolves to the side vacuum. The colored areas are the parameter ranges allowed by the latest observations of BICEP/$Keck$ collaboration \cite{BICEP:2021xfz}, where the color gradient from blue to red corresponds to $r$ increases from 0.001 to the observational upper limit 0.036. The dotted lines are the boundaries that $n_s$ exceeds the observational upper (pink line) or lower (green line) limit. The gray areas represent the parameter ranges with inadequate inflation, namely, the maximal $e$-folding number of inflation cannot reach $N=50$ or $60$.}
	\label{WR3para}
\end{figure}

Now we discuss the general cases with various $\zeta$ and $\gamma$ by numerical treatment. The results are shown in Fig.~\ref{WR3para}. Here the parameter ranges satisfying observational constraints (see blue area in Fig.~\ref{r_ns}) are marked with colored areas, where the color gradient from blue to red corresponds to ascending value of $r$. The gray areas represent that the potential defined by these parameters cannot support an adequate inflation. In other words, their maximal $e$-folding number is unable to reach $N=50$ or $60$. The white areas are the parameter ranges that can give rise to ample inflation, but their prediction of $n_s$ or $r$ has been excluded by the observation constraints. Here we mark two dotted lines to distinguish the boundaries of constraints. Beyond the pink one indicates a large $n_s$ that exceeds the observational upper limit, while beyond the green one signifies a too small prediction.

Let us focus on the colored parameter ranges that are allowed by observations. In the $|\zeta|\gg 1000$ case, the result is roughly equivalent to the analytical calculation shown in the last subsection. The prediction of $r$ is limited to $0.002<r<0.006$. However, distinctive situations appear when $|\zeta|$ is small. First, when $-1000<\zeta<-200$, the restrictions on $\gamma$ is relaxed, which can stand $|\gamma|\sim 6\times10^{-3}$ at most. Besides, the upper limit of $r$ is greatly expanded. There is even a small parameter range that gives $r>0.01$. We show an example as the green area in Fig.~\ref{r_ns}. It clearly shows a distinguishable feature from the Weyl $R^2$ model and the $R^3$-extended Starobinsky model. If the next generation experiment of CMB $B$-mode polarization detects the primordial gravitational waves with $r>0.01$, it may support Weyl $R^2+R^3$ model. Another notable feature emerges at $0<\zeta<200$, where the negative $\gamma$, even if very small, can greatly affect the predictions of primordial perturbations. Actually, there are some cases with small positive $\zeta$ and small negative $\gamma$ can give proper $r$ and $n_s$ that match the observation constraints, and generally, $r$ is extremely small. For instance, when $\zeta=80$, $\gamma=-4\times 10^{-8}$, and $N=60$, we have $n_s=0.963$ and $r=3\times10^{-4}$. 

\section{Inflation to the center}

As we mentioned earlier, the third vacuum appears at $\Phi=0$ in the case of $\zeta>0$ and $\gamma>0$, and if the initial value satisfies $|\Phi_i|<|\Phi_h|$ [$\Phi_h$ is defined in Eq.~(\ref{Phitop+})], inflation can happen in the evolution of $\Phi$ to 0. Actually, the situation is more complicated. A process called ``oscillating inflation'' \cite{Damour:1997cb, Liddle:1998pz, Taruya:1998cz, Cardenas:1999cw, Lee:1999pta, Sahni:1999qe, Tsujikawa:2000kw, Sami:2001zd, Dutta:2008px, Johnson:2008se, MohseniSadjadi:2013iou, Cembranos:2015oya, Goodarzi:2016iht} will continue immediately after the end of slow-roll inflation because the scalar potential in this case is a nonconvex function in the region close to the minimum, which means there is $\frac{\mathrm d^2V}{\mathrm d\Phi^2}<0$ when $\Phi$ nears 0. In other words, for such a nonconvex potential, despite the slow-roll conditions ($\epsilon\ll 1$ and $|\eta|\ll 1$) has been violated during the bottom oscillation of the inflationary potential, the universe can keep accelerating expansion until the average amplitude of the inflaton's oscillation becomes lower than the borderline of $\frac{\mathrm d^2V}{\mathrm d\Phi^2}$ from negative to positive (if there is a rounded transition in a small enough $\Delta\Phi$ at the bottom to connect the left and right sides of the potential, see \cite{Damour:1997cb}), or until the contribution of the radiation produced in reheating process becomes non-negligible.

\begin{figure}
	\centering
	\includegraphics[width=0.8\textwidth]{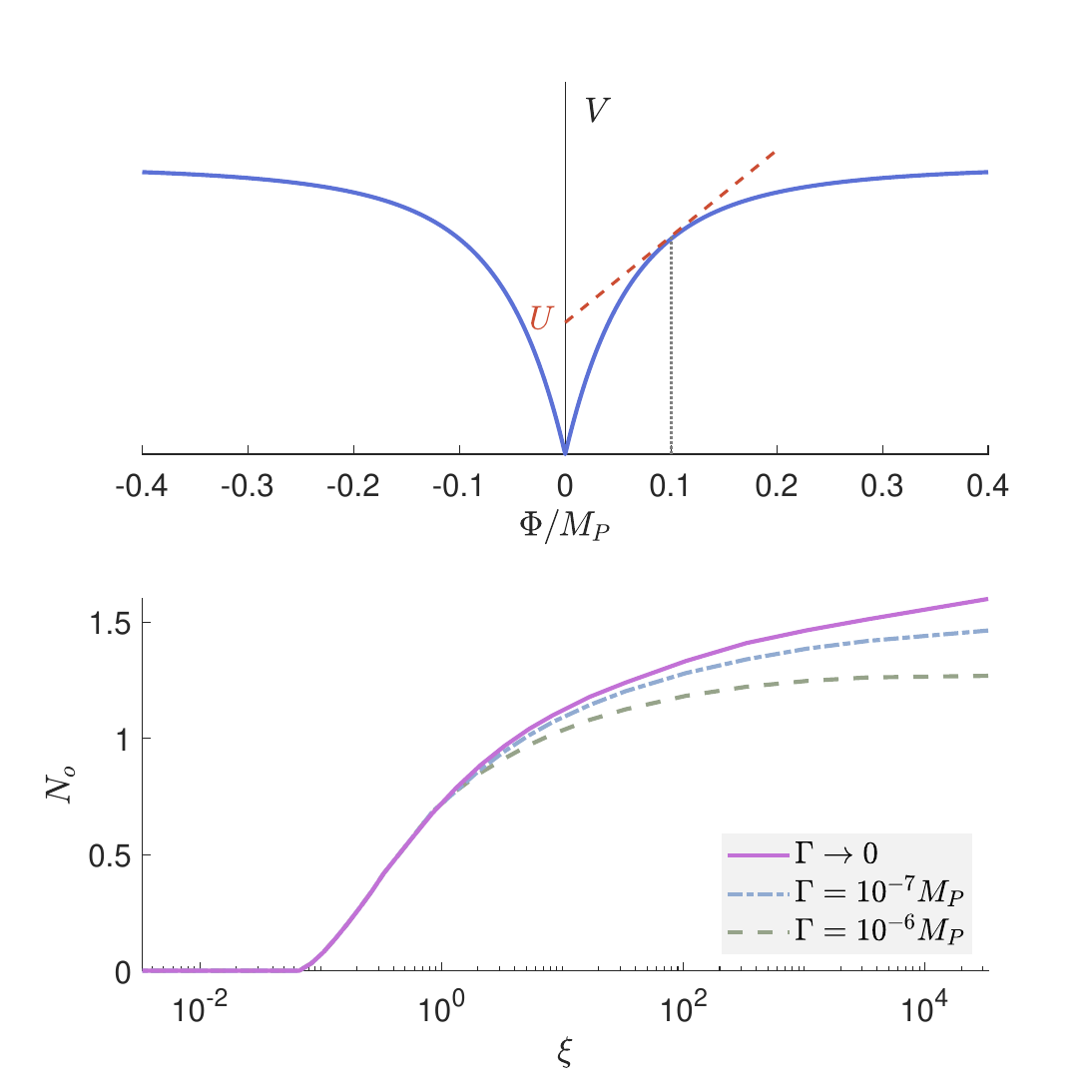}
	\caption{Oscillating inflation in the center-evolving pattern of Weyl $R^2+R^3$ model. The upper part is a diagram for visualizing the condition of oscillating inflation, where the effective equation of state $\langle w\rangle<-\frac{1}{3}$ is equated with that the intercept $U$ of the tangent to a certain point on the potential corresponding to the average amplitude is positive. The lower part shows the increased $e$-folding number during the oscillating inflation for various $\xi$ and reheating efficiency $\Gamma$.}
	\label{No}
\end{figure}

It is helpful to understand the behavior of oscillating inflation from the perspective of the effective equation of state. For an oscillating scalar field $\Phi$, its effective equation of state in one oscillating period is defined as
\begin{equation}
	\langle w\rangle\equiv\frac{\langle p\rangle}{\langle\rho\rangle}=\frac{\langle\dot\Phi^2-\rho\rangle}{\langle\rho\rangle}=\frac{\langle\dot\Phi^2\rangle}{V_m}-1=\frac{\langle\Phi\frac{\mathrm dV}{\mathrm d\Phi}\rangle}{V_m}-1=1-\frac{2\langle V\rangle}{V_m},
\end{equation}
where $\langle\rangle$ means the average value in one oscillation period, and $V_m$ represents the maximal potential of this oscillation period. The accelerating expansion of the universe requires $\langle w\rangle<-\frac{1}{3}$, which is equivalent to the following relation
\begin{equation}
	U\equiv\langle V-\Phi\frac{\mathrm dV}{\mathrm d\Phi}\rangle>0.
\end{equation}
In fact, $U$ amounts to the intercept of the tangent to the potential at a certain $\Phi$, shown as the upper part of Fig.~\ref{No}. As long as the intercept is positive and the contribution of radiation is insignificant, the accelerating expansion will proceed successfully. This is the reason why a nonconvex potential can bring about oscillating inflation.

For the process with oscillating inflation, the definition of $e$-folding number should be replaced to
\begin{equation}
	\tilde N\equiv \ln\frac{a_fH_f}{a_iH_i}\equiv\ln\frac{a_eH_e}{a_iH_i}+\ln\frac{a_oH_o}{a_iH_i}\simeq N+N_o,
	\label{Ntilde}
\end{equation}
where the subscripts $i$ and $e$ have been defined in the last section, $a_f$ and $H_f$ represent the cosmic scale and Hubble parameter when the full inflationary period ends, $a_o$ and $H_o$ represent their multiple of increase or decrease during the oscillating inflation. It indicates that the new definition is equivalent to adding a correction $N_o$ based on the $e$-folding number of slow-rolling period if we take $H_e\approx H_i$. Generally, $N_o$ is related to the shape of potential near its minimum, reheating efficiency, and the scale of the aforementioned rounded bottom. We have discussed in Appendix \ref{ap1} that a higher-order $\hat R^4$ term can bring our model a rounded bottom. But here we consider this term is small enough for simplicity, that is, $N_o$ depends only on the first two aspects. For the shape of potential, actually, our model has the following approximate form near the center minimum
\begin{equation}
	V(\Phi)\simeq-\frac{\xi\Phi^2}{2\alpha}+\frac{\xi^2\Phi^4}{3\alpha}\left[\left(1+\frac{1}{\xi\Phi^2}\right)^{3/2}-1\right],
	\label{V0}
\end{equation}
where $\xi\equiv\frac{\alpha^2}{3\beta\zeta}$. Since $\alpha$ determines the height of the potential, which has been fixed for each set of $\zeta$ and $\beta$ according to the observation result of $\Delta^2_s\sim \frac{V}{24\pi^2\epsilon}\sim2.1\times10^{-9}$ \cite{Planck:2018vyg}, the shape of the potential is essentially determined by $\xi$ in the oscillatory region. For reheating efficiency, we consider a constant transfer rate $\Gamma$ and the transferred energy all turns to radiation $\rho_r$
\begin{equation}
	\ddot\Phi+(3H+\Gamma)\dot\Phi+\frac{\mathrm dV}{\mathrm d\Phi}=0,
\end{equation}
\begin{equation}
	\dot\rho_r+4H\rho_r-\Gamma\dot\Phi^2=0.
\end{equation}
Then $N_o$ is substantially related to the parameters $\xi$ and $\Gamma$.

We numerically solve the above equations, and visualize in the lower part of Fig.~\ref{No}. It is transparent that if $\xi\gg 0.1$, oscillating inflation will bring appreciable correction to the $e$-folding number. Because an inefficient reheating process will postpone the end of the oscillating inflation, we can see a smaller $\Gamma$ corresponds to a larger $N_o$ for a certain $\xi$. However, $N_o$ will tend to a fixed value as $\Gamma$ decreases. This property can be understood as follows. We can prove that the potential has a quasilinear form when $\Phi\rightarrow0$
\begin{equation}
	V|_{\Phi\rightarrow0}\simeq\frac{\sqrt{\xi}}{3\alpha}|\Phi|,
\end{equation}
which implies that $U|_{\Phi\rightarrow0}\rightarrow 0$ according to its definition as the intercept of the tangent to the potential. Hence $\langle w\rangle$ will quickly converge to $-\frac{1}{3}$ as the oscillation proceeds, and $N_o$ will soon grow to a nearly constant maximum if $\Gamma$ is too small to make the universe promptly produce enough radiation to stop the oscillating inflation. This is the reason why $N_o$ has an extreme for each $\xi$.

\begin{figure}
	\centering
	\includegraphics[width=0.8\textwidth]{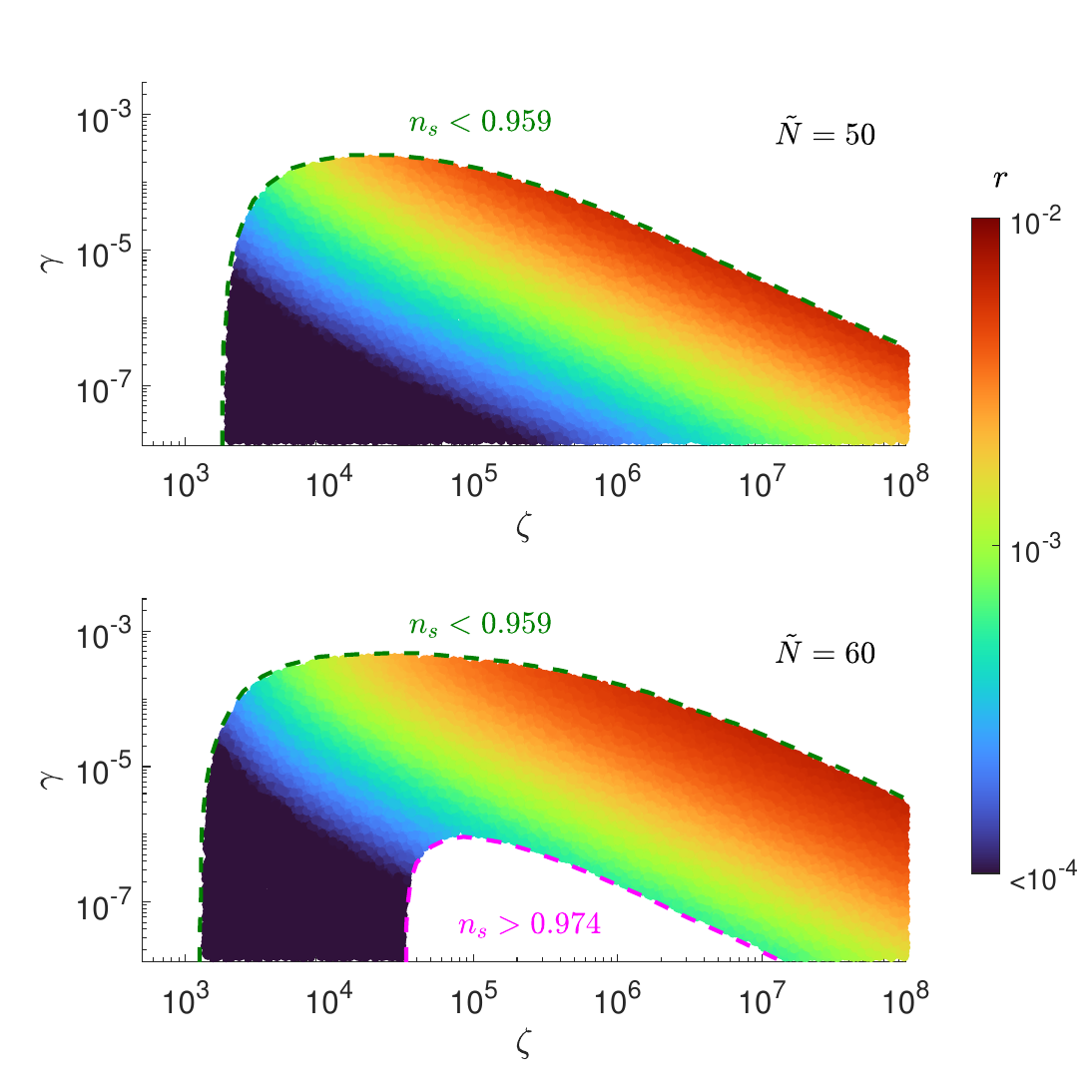}
	\caption{Possible parameter space for Weyl $R^2+R^3$ model when $\Phi$ evolves to the center vacuum. Here the total $e$-folding number $\tilde N\equiv N+N_o$ is considered with $\Gamma\rightarrow0$. The meaning of markers is the same as that in Fig.~\ref{WR3para}, except for the color correspondence of $r$.}
	\label{WR3to0}
\end{figure}

Now we consider the reheating is inefficient, that is to adopt $N_o$ with $\Gamma\rightarrow 0$, to derive the slow-roll $e$-folding number $N$ corresponding to $\tilde N\sim(50,60)$, and then to calculate $n_s$ and $r$ for various parameters $\zeta$ and $\gamma$. The viable parameter space is depicted in Fig.~\ref{WR3to0}, where the meaning of markers is the same as that in Fig.~\ref{WR3para}, except for the scale of color bar. It is evident that the observation constraint on $n_s$ limits the parameters to $\zeta>10^3$ and $\gamma<5\times10^{-4}$. $r$ has an upper limit $\sim 0.006$, but no lower limit in this case.

\section{Conclusions}

Cosmological observations have suggested that our universe has a nearly scaling invariant power spectrum of the primordial density perturbation, which motivates the scaling symmetry as the possible feature of the underlying fundamental theories that lead to inflation. We present the theoretical formalism of the Weyl scaling invariant gravity, $\hat{R}^2+\hat{R}^3$. We show this model in Eq.~(\ref{LWR3}) can be rewritten equivalently to the Einstein gravity coupled with a massive gauge boson, and a scalar field as the inflaton. We further discuss the viable ranges of the scalar potential according to the requirement for reality and demonstrate how the $R^3$ term would affect the shape of potentials. Compared with the Weyl $R^2$ inflationary potential~\cite{Tang:2020ovf,Wang:2022ojc} with two side minima, the $R^3$ extension brings an additional minimum at center. Hence, there are two viable scenarios for the inflation in this model. The first is to roll toward the side minima, while the other is a new situation of rolling toward the center minimum. Both scenarios allows viable parameter spaces that be probed by future experiments on cosmic microwave background and primordial gravitational wave.

For the first scenario, we calculate the spectral index $n_s$ and tensor-to-scalar ratio $r$ of primordial perturbations both analytically and numerically, and contrast the parameter spaces with the latest observational constraints. The results manifest that the level of cubic curvature is limited to $|\gamma|<6\times10^{-3}$, and the prediction of $r$ in this pattern has a wide range from $\mathcal O(10^{-4})$ to the upper limit of the observations, $\mathcal O(10^{-2})$. These results are significantly different from the $R^3$-extended Starobinsky model.

For the second scenario, a special process called oscillating inflation emerges after the familiar slow-roll inflation because the potential near the center minimum is a nonconvex function that can lead to a sufficiently negative value of average equation of state. We calculate the correction of $e$-folding number in the oscillating inflation stage, and then derive the viable parameter spaces. The results indicate that the parameters are limited to $\gamma<5\times10^{-4}$ and $\zeta>10^3$. Moreover, $r$ has an upper limit $\sim 0.006$, but no lower limit. 

\begin{acknowledgments}
Q.Y.W. and Y.T. thank Shi Pi for helpful discussions. Y.T. is supported by National Key Research and Development Program of China (Grant No.2021YFC2201901), and Natural Science Foundation of China (NSFC) under Grants No.~11851302. Y.L.W. is supported by the National Key Research and Development Program of China under Grant No.2020YFC2201501, and NSFC under Grants No.~11690022, No.~11747601, No.~12147103, and the Strategic Priority Research Program of the Chinese Academy of Sciences under Grant No. XDB23030100.		
\end{acknowledgments}

\appendix
\section{Effect of an extra $\hat R^4/\varphi^4$ term}\label{ap1}

As we mentioned in the introduction, any high-order curvature terms may exist and have an effect on the inflationary potential from the viewpoint of effective field theory. Therefore it is instructive to inspect how an extra tiny $\hat R^4$ term affects our model. We expand Eq.~(\ref{FR}) to the following form
\begin{equation}
	F(\hat R,\varphi)=\varphi^2\hat R+\alpha \hat R^2+\frac{\beta}{\varphi^2}\hat R^3+\frac{\delta}{\varphi^4}\hat R^4.
\end{equation}
Then the frame fixing equation is rewritten as
\begin{equation}
	\varphi^2+2\alpha\chi^2+ 3\beta\chi^4/\varphi^2+4\delta\chi^6/\varphi^4=1.
\end{equation}
It is a cubic equation for $\chi^2$, which has three roots as the following form corresponding to $n=1,2,3$ respectively,
\begin{equation}
	\chi^2=-\frac{\beta\varphi^2}{4\delta}+\omega^{n}\left[\Lambda+\sqrt{\Delta}\right]^{1/3}+\omega^{2n}\left[\Lambda-\sqrt{\Delta}\right]^{1/3},~n=1,2,3,
	\label{root3}
\end{equation}
where the phase factor $\omega=\frac{-1+\sqrt3 i}{2}$, and
\begin{equation} 	
	\Lambda=\frac{(4\alpha\beta\delta-\beta^3-8\delta^2)\varphi^6+8\delta^2\varphi^4}{64\delta^3},
\end{equation}
\begin{equation} 	
	\Delta=\frac{\varphi^{12}}{6912\delta^4}\left[32\alpha^3\delta-9\alpha^2\beta^2-(108\alpha\beta\delta-27\beta^3)\frac{\varphi^2-1}{\varphi^2}+108\delta^2\frac{(\varphi^2-1)^2}{\varphi^4}\right].
	\label{Delta}
\end{equation}
Only one of the three roots can return to Eq.~(\ref{root2}) under $\delta\rightarrow0$ limit, which corresponds to a correct physical situation. We call it a \emph{proper root}. Which root is the proper root depends on the sign of $\beta$ and $\delta$. When $\beta>0$ and $\delta>0$, the proper root is Eq.~(\ref{root3}) with $n=3$. When $\beta<0$ and $\delta>0$, it is $n=1$. And when $\delta<0$, it is $n=2$. Furthermore, note that Eq.~(\ref{Delta}) is the discriminant of the cubic equation. Only when $\Delta\leq0$, the proper root is a real root that is physically allowed. This actually constrains the upper limit of the parameter $\delta$ in some cases.

\begin{figure}
	\centering
	\includegraphics[width=0.75\textwidth]{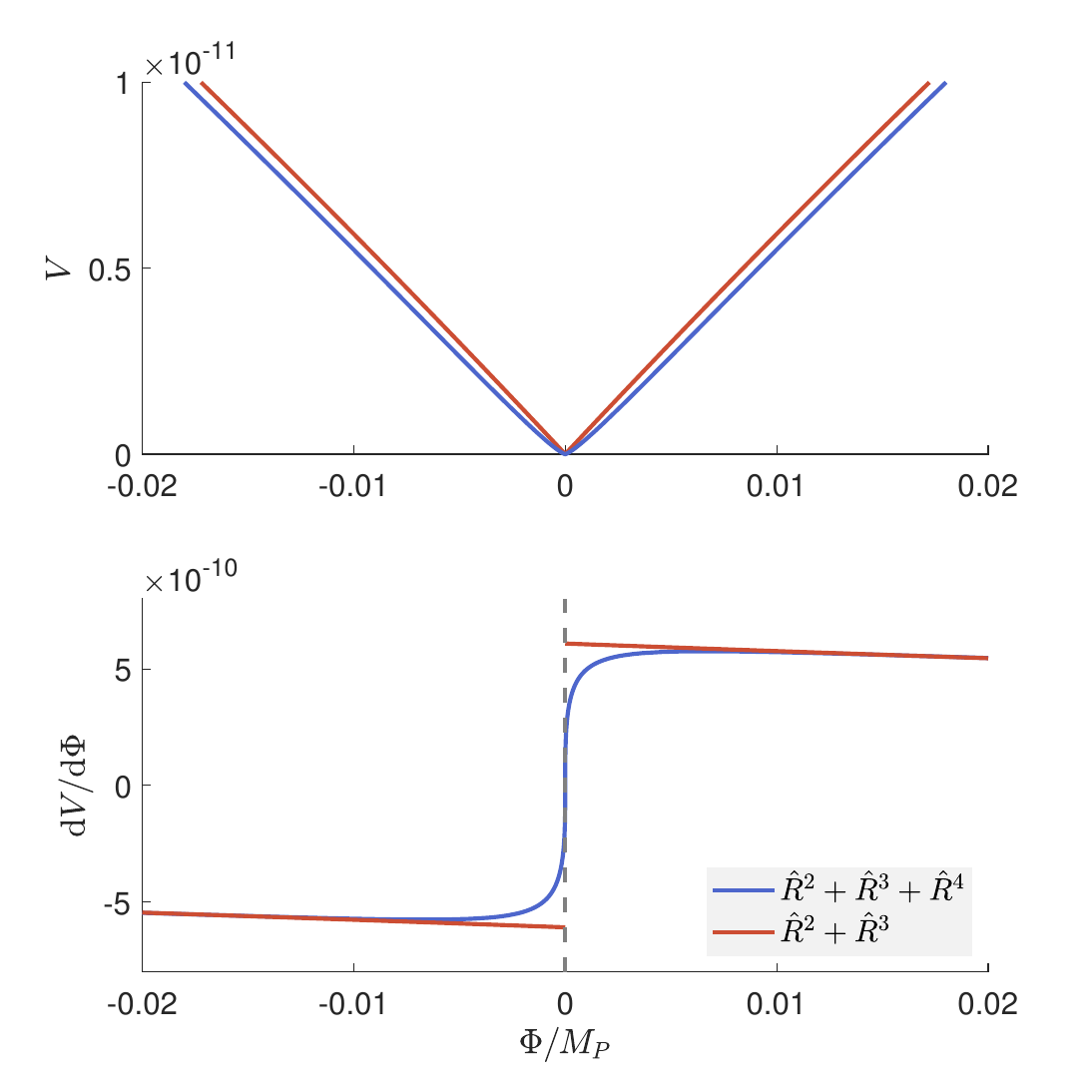}
	\caption{Effective scalar potentials and derivatives of Weyl $R^2+R^3$ model (red) and Weyl $R^2+R^3+R^4$ model (blue) near the center minimum with $\alpha=10^9$, $\beta=10^{-4}\alpha^2$, and $\delta=10^{-10}\alpha^3$. The $\hat R^4$ term brings a rounded bottom to the potential when $\delta>0$.}
	\label{R4}
\end{figure}

Using the same method as Sec.~II.A., we can derive the scalar potential as
\begin{equation}
	\begin{aligned}
		V(\varphi) =\frac{\alpha}{2}\chi^4+\frac{\beta}{\varphi^2}\chi^6+\frac{3\delta}{2\varphi^4}\chi^8.
	\end{aligned}
\end{equation}
We find that when $\delta$ is extremely small, the $\hat R^4$ term basically does not affect the shape of the potential. Its impact only concentrates around $\varphi^2=0$ (or $\Phi=0$). Figure~\ref{R4} shows the potential and its derivative near the central minimum with $\beta>0$ and $\delta>0$, here we have introduced Eq.~(\ref{Phi}). We are surprised to find that the $\hat R^4$ term eliminates the nonanalytic point of the Weyl $R^2+R^3$ model, which may have caused problems. The scale of the rounded bottom is proportional to the parameter $\delta$. Thus when $\delta$ is small, the physical quantities of slow-roll inflation (e.g., $n_s$ and $r$) will not be affected.

\section{Analytical treatment of Starobinsky inflation}\label{ap2}

We give an analytical calculation of the tensor-to-scalar ratio $r$ and spectral index $n_s$ in the Starobinsky inflationary model, namely, the Einstein gravity modified by a $R^2$ term. The effective scalar potential can be written as
\begin{equation}
	V(\phi)=\frac{1}{8\alpha}\left(1-e^{-\sqrt{2/3}\phi}\right)^2,
	\label{VStaro}
\end{equation}
where $\alpha$ is the coefficient of $R^2$. The relevant two slow-roll parameters are computed as
\begin{equation}
	\epsilon=\frac{4}{3}\frac{1}{\left(e^{\sqrt{2/3}\phi}-1\right)^2},~~~\eta=-\frac{4}{3}\frac{e^{\sqrt{2/3}\phi}-2}{\left(e^{\sqrt{2/3}\phi}-1\right)^2}.
	\label{enS}
\end{equation}
Since inflation ends when $\epsilon\sim 1$ is reached first $(\eta \simeq -0.15)$, we have
\begin{equation}
	\phi_e=\sqrt{\frac{3}{2}}\ln\left(1+\frac{2}{\sqrt3}\right)\simeq 0.94M_P.
\end{equation}
Then according to Eq.~(\ref{N}), the $e$-folding number is
\begin{equation}
	\begin{aligned}
		N=\left[\frac{3}{4}\left(e^{\sqrt{2/3}\phi}-\sqrt{\frac{2}{3}}\phi\right)\right]_{\phi_i}^{\phi_e}=\frac{3}{4}\left[e^{\sqrt{2/3}\phi_i}-e^{\sqrt{2/3}\phi_e}-\sqrt{\frac{2}{3}}(\phi_i-\phi_e)\right].
	\end{aligned}
\end{equation}
For $N\sim(50,60)$, we find that approximately 
\begin{equation}
	\phi_i\simeq\sqrt{\frac{3}{2}}\ln\left[\frac{4}{3}(N+4.3)\right].
\end{equation}
Substituting it into Eq.~(\ref{enS}), we finally derive
\begin{equation}
	r=16\epsilon=\frac{12}{(N+3.55)^2},
\end{equation}
\begin{equation}
	n_s=1-6\epsilon+2\eta=1-\frac{2}{N+3.55}-\frac{3}{(N+3.55)^2}.
\end{equation}
These results are shown as the red line in Fig.~\ref{r_ns}.


\end{document}